\documentclass[preprint,showkeys,aps,amssymb,prd]{revtex4}

\usepackage{bm,graphicx, wrapfig}
\usepackage{amsmath,amssymb,latexsym}

\newcommand\beq{\begin{equation}}
\newcommand\beqa{\begin{eqnarray}}
\newcommand\beqan{\begin{eqnarray*}}
\newcommand\eeq{\end{equation}}
\newcommand\eeqa{\end{eqnarray}}
\newcommand\eeqan{\end{eqnarray*}}

\newcommand\mdot{{\sf m}_\bullet}

\newcommand\alphahat{\hat{\alpha}}

\begin{document}

\title{Strong and Weak Deflection of Light in the Equatorial Plane of a Kerr Black Hole} 

\author{S. V. Iyer}
\email{iyer@geneseo.edu}
\author{E. C. Hansen}
\email{ech3@geneseo.edu}
\affiliation{Department of Physics \& Astronomy, State University of New York at Geneseo,\\
1 College Circle, Geneseo, NY 14454.}

\begin{abstract}

Analytical series expansions for the bending of light in the equatorial plane of a Kerr black hole are presented
in both the strong and weak deflection regimes.  It is critical that these are known in analytical form so further
analysis can be done for predicting different properties of images formed in gravitational lensing.  Starting with the exact
bending angle in terms of the spin parameter, we apply a perturbative scheme for rewriting the 
bending angle as series expansions in terms of the impact parameter 
of the incident light ray.  The asymmetry introduced by the black hole spin results in spin-dependent shifts in 
image positions.  We apply our results for the case of a galactic supermassive black hole to 
predict angular shifts of relativistic images from the optic axis.  This would not be possible without the perturbative
expansions in the strong deflection regime, only in which relativistic images have a chance of 
being resolved by future telescopes.

\end{abstract}


\maketitle

\section{Introduction}
\label{sec:intro}

Observations of the bending of light near the sun in the early twentieth century and Einstein rings 
in deep space Hubble space telescope images in the last decade or so are both examples of tests of general relativity in the 
weak field limit, or more precisely, the weak deflection limit.  As the light ray skims by, not far from the 
horizon, however, a richer variety of lensing effects are predicted to occur, awaiting observation by telescopes
with much higher viewing power than what is available today.  Bending angle 
calculations \cite{darwin, atkinson, luminet, ohanian, chandra} for Schwarzschild and Kerr geometries 
show that as we approach the depths of the gravitational potential, multiple looping of the light ray around the
center of attraction is possible resulting in what are known as relativistic images (see for example, \cite{MTW73}).  
In the case of a Schwarzschild black hole, as the distance of closest approach $r_0$ nears the critical value 
of $3\mdot$, where $\mdot=G M/c^2$ is its gravitational radius, the bending angle has a logarithmic form.  
When expressed in terms of the invariant quantity $b'=1-b_c/b$, where $b_c=3\sqrt{3}\mdot$ this strong
deflection limit (SDL) of the bending angle is given by \cite{iyerpetters}
\beq
\label{AlphaSStrong}
\alphahat (b')= - \pi+\log\left[\frac{216\, (7 - 4 \sqrt{3})}{b'} \right] +  \mathcal{O}[b']+...
\eeq
This result is similar, but not identical, to the Darwin logarithmic term \cite{darwin}; this result takes into account 
an improvement to Darwin's result close to the photon sphere.  The detailed formulation that makes this
modification to the strong deflection limit was recently published in \cite{iyerpetters}.  

In this paper, we show that the analysis can be applied to the spinning, or Kerr, black hole 
for the case when the light ray stays on the equatorial plane.  In order to determine the
perturbative series for the bending angle, an explicit expression in terms of just the 
spin and mass of the black hole as a function of the impact parameter is needed as the starting point.
Such an expression was recently derived in \cite{iyerhansen1}.  The bending angle was expressed 
explicitly in terms of the black hole mass $\mdot=GM/c^2$ and the black hole spin parameter $a=J/Mc$, where $J/M$ is the 
angular momentum per unit mass of the black hole.  With a new definition for $b'$ that 
includes the non-zero spin of the black hole, we define precisely the approach
to the critical impact parameter, and thus the strong deflection limit (SDL).  

The weak deflection limit (WDL), on the other hand, in both the Schwarzschild and the Kerr case are 
easily defined in terms of the impact parameter as the limit $b\rightarrow\infty$.  In this 
limit, the weak deflection bending angle series for the Schwarzschild case is given by
\beq
\label{AlphaSWeak} 
\alphahat (b) =4\left(\frac{\mdot}{b}\right)+\frac{15\pi}{4}\left(\frac{\mdot}{b}\right)^2+
\mathcal{O}\left[\left(\frac{\mdot}{b}\right)^3\right]+...
\eeq
A partial generalization of this result for the Kerr case in terms of the coordinate-dependent
variable $r_0$ was obtained by Boyer-Lindquist, Skrotskii, Plebanski \cite{boy-lind, 
skrotskii, plebanski} and from a numerical treatment by Rauch and Blandford \cite{rauch-blandford}.
The Kerr WDL series in terms of the impact parameter was recently obtained 
in a completely different context by Petters \cite{petters}.  

In this paper, we continue to work in the same perturbative framework that was started in
\cite{iyerpetters} and apply it to the case of deflection of rays confined to the 
Kerr equatorial plane.  Bending angles, and therefore the position and magnification 
of images, depend crucially on whether the light ray is traversing in the same or opposite 
direction to the rotation.  For this reason, throughout the analysis, we carefully keep track of 
whether the ray orbits are direct or retrograde.  The perturbative corrections are obtained 
in a manner very similar to that for the Schwarzschild case presented in \cite{iyerpetters}.  

The need for analytical results for the bending angle cannot be overemphasized given that 
important lensing variables like image positions, magnifications and time delays
depend crucially on these results.  See \cite{igorbray}-\cite{vazquez-esteban} for many different 
approaches towards analytical results. Perturbative analysis is extremely useful in order to 
verify, at least to leading orders, the predictions of strong deflection lensing.  Specific to 
the Kerr case, we also need to be able to glean the contribution to light deflection arising purely from 
spin.  Furthermore, higher order relativistic images can be analyzed only in the strong
deflection regime, making this an important approach to test general relativity 
beyond the usual classical tests in the weak field regime.

In Section 2, we start with the bending angle result from \cite{iyerhansen1} and definitions of some 
of the variables to be used in the series expansions.  We present the weak and strong deflection series
expansion terms in Sections 3 and 4 respectively, along with numerical plot comparisons of 
these terms with the exact result.  

\begin{figure}[htp] 
\begin{center} 
\includegraphics[width=3.3in]{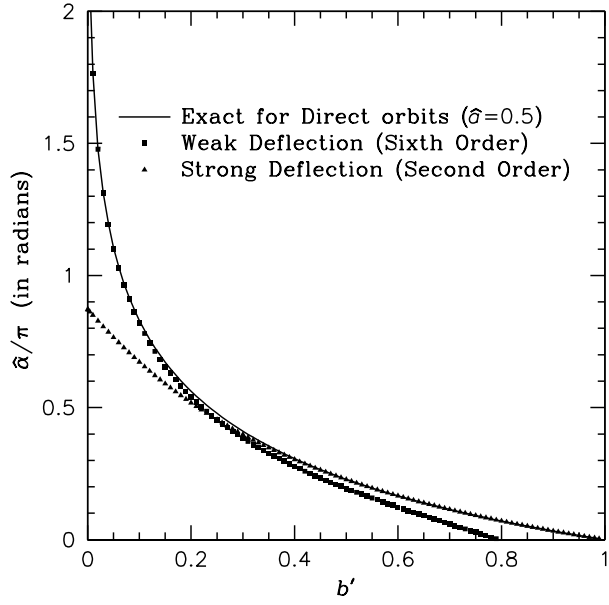}
\caption{\small \sl The 2nd-order strong deflection and the 6th-order weak deflection are plotted alongside
the numerically integrated exact formal Kerr bending angle $\alphahat$ (in units of $\pi$) for {\em direct} orbits.} 
\label{BestDp5} 
\end{center} 
\end{figure} 

\begin{figure}[htp] 
\begin{center} 
\includegraphics[width=3.3in]{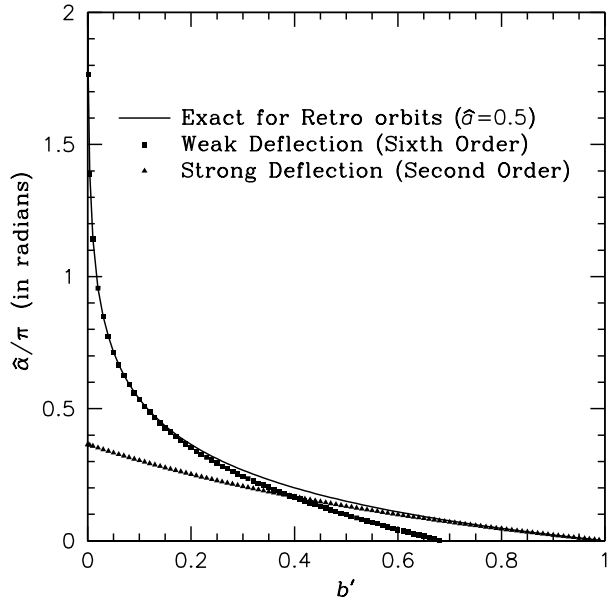}
\caption{\small \sl The 2nd-order strong deflection and the 6th-order weak deflection are plotted alongside
the numerically integrated exact formal Kerr bending angle $\alphahat$ (in units of $\pi$) for {\em retro} orbits.} 
\label{BestRp5} 
\end{center} 
\end{figure} 

As a quick preview, we present here two plots that illustrate the main focus of this paper: the 2nd-order 
strong deflection and the 6th-order weak deflection bending angles are plotted alongside the exact bending 
angle for comparison, for direct and retrograde orbits in 
Figures \ref{BestDp5} and \ref{BestRp5} showing that our series expansions have excellent accuracy; the SDL 
series and WDL series apply nicely in the two regimes. Note
that the two assumptions we have made here are: (1) the light ray stays on the equatorial plane, and 
(2) the spin parameter stays in the range $0\le a < \mdot$.  As can be seen in the plots, the agreement 
between the exact and our series expansions is better in the direct than in the retro 
case.  

In section 5 is presented an application of both series to the simplest
case of lensing geometry in which the source, lens and the observer are in perfect alignment.  We 
show that the usual calculation for the Einstein angle that uses the WDL
bending angle is applicable in the Kerr case as well.  However, since there is no contribution to first order,
the Einstein ring radius and angle are indistinguishable from the static case.  In the strong 
deflection limit, on the other hand, the effect of spin on the angular positions 
relativistic images is relatively much greater.  This numerical calculation is analogous to the one for
the Schwarzschild case found in Virbhadra and Ellis \cite{vir-ellis}.  We show that the shift in the 
image position increases with the black hole angular momentum.  

\section{Formal Exact Bending Angle for the Equatorial Kerr case}

Let us now consider a light ray that starts in the 
asymptotic region and approaches the black hole, with $r_0$ as the distance of closest 
approach.  It then emerges and reaches an observer 
who is also in an asymptotic region.  

We will use the following convenient notation:
\beq
h=\frac{\mdot}{r_0}\qquad\omega_s=\frac{a}{b_s} \qquad
{\rm and}\qquad\omega_0=\frac{a^2}{\mdot^2}
\eeq
where $b_s=sb$ is the invariant impact parameter.  The parameter $s$ will be used to keep
track of the sign of the impact parameter relative to the black hole spin.  : $s=+1$ for direct 
orbits and $s=-1$ for retrograde orbits.  Quantities that have $s$ as subscript obey the same sign
convention.  For example, $\omega_s$ takes on the appropriate sign for direct and 
retrograde orbit.  Note that in the limit \{$\omega_s, \omega_0 \rightarrow 0$\}, we
recover the zero-spin Schwarzschild case, and in the limit $h\rightarrow 0$, we have the 
zero-deflection flat metric limit.   With $a=J/Mc$ it is also convenient to introduce 
\beq
\hat{a}=\frac{a}{\mdot}=\frac{Jc}{GM^2},
\eeq
as the ``normalized" spin parameter.  We will limit ourselves to cases 
where $0\le \hat{a} \le 1$, with $\hat{a}=0$ being the Schwarzschild limit 
and $\hat{a}=1$ being extreme Kerr.  Next, we define critical 
parameters analogous to the Schwarzschild case in \cite{iyerpetters}:
\beq
h_{sc}=\frac{1+\omega_s}{1-\omega_s}  \qquad
{\rm and}\qquad r_{sc}=\frac{3\mdot}{h_{sc}}
\eeq
We also define the variable
\beq
h'=1-\frac{3h}{h_{sc}}\equiv1-3\left(\frac{\mdot}{r_0}\right)\left(\frac{1-\omega_s}{1+\omega_s}\right)
\eeq
with
\beq
1-3\left(\frac{\mdot}{r_0}\right)\left(\frac{1-\omega_s}{1+\omega_s}\right)\quad 
\xrightarrow{\quad a \rightarrow 0\quad} \quad1-\frac{3\mdot}{r_0}. \nonumber
\eeq

We have introduced these different quantities for the Kerr case, keeping in mind that they should go over 
to those defined in the Schwarzschild case smoothly when $a$ is set equal to zero.  So, as shown above, 
as $a\rightarrow 0$, $h_{sc}\rightarrow 1$ and we recover the definition of $h'$ in \cite{iyerpetters}.  In both
cases, $h\rightarrow 0$ at critical, and $h\rightarrow 1$ as $r_0$ approaches infinity.

\begin{figure}[htp] 
\begin{center} 
\includegraphics[width=3.3in]{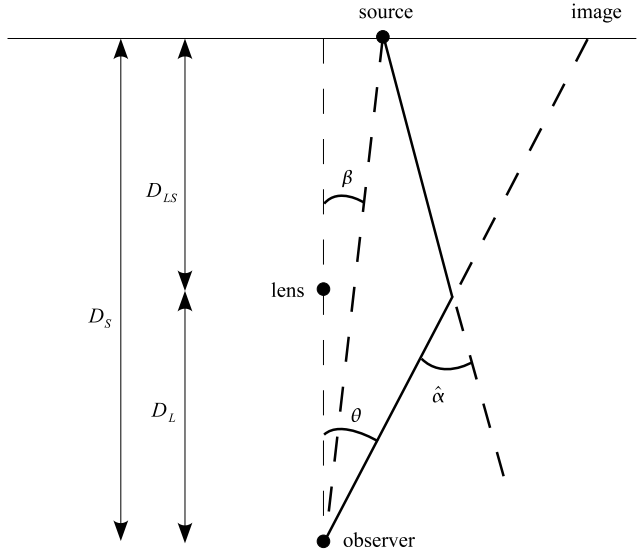}
\caption{\small \sl Thin lens geometry.} 
\label{LensGeom} 
\end{center} 
\end{figure} 

Critical values for the radius and the impact parameters in Kerr geometry are given by:
\beq \label{rcrit}
r_{sc}=3 \mdot \frac{\left(1-\dfrac{a}{b_{sc}}\right)}
{\left(1+\dfrac{a}{b_{sc}}\right)}
\eeq
and 
\beq \label{bcrit}
{(b_{sc}+a)}^3=27 \mdot^2 (b_{sc}-a).
\eeq
From a lensing perspective, we are interested in impact parameters just beyond the 
critical value (SDL) extending all the way to infinity (WDL).  We define the dimensionless 
quantity $b'$ as 
\beq 
b'=1-\frac{s b_{sc}}{b}
\eeq
where the insertion of the quantity $s$ guarantees that the $b'$ stays between $0$ and $1$.  Note that 
this definition goes over naturally in the Schwarzschild limit:
\beq
1-\frac{s b_{sc}}{b}\quad \xrightarrow{\quad a \rightarrow 0\quad} \quad1-\frac{3\sqrt{3}\mdot}{b}\nonumber
\eeq
As will become apparent in later sections, the ``normalized" impact parameter $b'$ is 
extremely convenient and natural for describing the range of values from critical all the way to infinity.

The lensing geometry is shown in Figure \ref{LensGeom} where
it is assumed as usual that the thickness of the lens plane is much smaller than the distances between
source and lens $D_{LS}$.  The distance (from the observer) to the lens and source are denoted 
by $D_{L}$ and $D_{S}$ respectively.  The angular position of the source, $\beta$ and the image, $\theta$
are also shown in the figure.

The deflection of the light ray from its original path $\alphahat$ is given by the following expression\cite{iyerhansen1}.
\begin{widetext}
\beq
\label{ExactKAlpha}
\alphahat=-\pi+\frac{4}{1-\omega_s}\sqrt{\frac{r_0}{Q}} 
\bigg\lbrace\Omega_+ \left[\Pi(n_+,k)
-\Pi(n_+,\psi,k)\right]+\Omega_- \left[\Pi(n_-,k)-\Pi(n_-,\psi,k)
\right]\bigg\rbrace, 
\eeq
\end{widetext}
where $\Pi(n_\pm,k)$ and $\Pi(n_\pm,\psi,k)$ 
are the complete and the incomplete elliptic integrals 
of the third kind respectively.
The argument $k^2$ is defined through the elliptic
integral as usual in the range $0\le k^2 \le 1$.  Note that in some references the variable 
is referred to as $k^2$ and in others simply as $k$.  The order in which the
arguments appear in $\Pi(n,\psi,k)$ also varies between different references 
and in Mathematica.

\noindent {\it Remark:} In Mathematica, the built-in mathematical
function for the incomplete elliptic integral of the third kind 
${\rm EllipticPi}[{\sf n},\phi,{\sf m}]$ is defined by
\beq
\int_0^\phi
\left[1 - {\sf n}\sin^2 \theta\right]^{-1}\left[1 - {\sf m} \sin^2 \theta\right]^{-1/2}d \theta \nonumber
\eeq
and the complete elliptic integral of the third kind is $ {\rm EllipticPi}[{\sf n, m}] = {\rm EllipticPi}[{\sf n},\pi/2, {\sf m}]$.

In the limiting case when $a \rightarrow 0$, we have $\Omega_+ =1$, 
$\Omega_-=0$ and $n_+=0$ we have the Schwarzschild result
\beqa
\alphahat&=&-\pi+4\sqrt{\frac{r_0}{Q}}\left[\Pi(0,k)
-\Pi(\psi,0,k)\right] \nonumber \\
&=& -\pi+4\sqrt{\frac{r_0}{Q}}\left[K(k)-F(\psi,k)\right],
\eeqa
where $K(k)$ and $F(\psi,k)$ are the complete and 
incomplete integrals of the first kind respectively.  In addition, in the limit 
when $\mdot=0$ (i.e., $h\rightarrow 0$) we recover zero deflection as expected.

The different variables are defined as follows:

\beq
\frac{r_0}{Q}=\frac{1}{h_{sc}\sqrt{\left(1-\dfrac{2h}{h_{sc}}\right)
\left(1+\dfrac{6h}{h_{sc}}\right)}}
\eeq
\beq
k^2=\frac{\sqrt{\left(1-\dfrac{2h}{h_{sc}}\right)
\left(1+\dfrac{6h}{h_{sc}}\right)}
+\dfrac{6h}{h_{sc}}-1}{2\sqrt{\left(1-\dfrac{2h}{h_{sc}}\right
)\left(1+\dfrac{6h}{h_{sc}}\right)}}
\eeq
\beq
\psi=\arcsin{\sqrt{\frac{1-\dfrac{2h}{h_{sc}}-\sqrt{\left(1-\dfrac{2h}{h_{sc}}\right)
\left(1+\dfrac{6h}{h_{sc}}\right)}}{1-\dfrac{6h}{h_{sc}}
-\sqrt{\left(1-\dfrac{2h}{h_{sc}}\right)\left(1+\dfrac{6h}{h_{sc}}\right)}}}}
\eeq
\begin{widetext}
\beq
\Omega_\pm=\frac{\pm (1\pm\sqrt{1-\omega_0})
(1-\omega_s)\mp\omega_0/2}{\sqrt{1-\omega_0}\left( 1\pm\sqrt{1-\omega_0}
-\dfrac{\omega_0 h_{sc}}{4}\left[1-\dfrac{2h}{h_{sc}}-\sqrt{\left(1-\dfrac{2h}{h_{sc}}\right)
\left(1+\dfrac{6h}{h_{sc}}\right)}\;\right]\right)}
\eeq
\beq
n_\pm=\frac{1-\dfrac{6h}{h_{sc}}-\sqrt{\left(1-\dfrac{2h}{h_{sc}}\right)
\left(1+\dfrac{6h}{h_{sc}}\right)}}
{1-\dfrac{2h}{h_{sc}}-\sqrt{\left(1-\dfrac{2h}{h_{sc}}\right)
\left(1+\dfrac{6h}{h_{sc}}\right)}-\dfrac{4}{\omega_0 h_{sc}}
\left(1\pm\sqrt{1-\omega_0}\right)}
\eeq
\end{widetext}

We note here that the quantity $b'$ appears in this expression via $r_0,h, h_{sc}$, and 
$\omega_s$, while $\omega_0=a^2/\mdot^2$ is independent of the impact parameter.  Any 
quantity that has an ``$s$" in the subscript takes on a negative sign 
when on the retro side.  The angle $\alphahat$ itself stays positive, i.e., the light ray is still deflected
towards the axis of rotation albeit to a lesser extent on the retro side \cite{iyerhansen1}.  In the 
next two sections, we outline the series expansions for the weak and
strong deflection limits.

\section{Expansion of the bending angle beyond critical}

As we did in \cite{iyerpetters}, we show that the SDL bending angle can be expressed as 
an ``affine perturbation'' series in $b'$.  The quantity $b'$ is now defined appropriately for 
the equatorial Kerr case.  The choice of variables turns out to be very important for carrying
out the entire calculation successfully.  Moreover, the series terms in the equatorial Kerr case
become much lengthier because of the contribution from the non-zero spin. 

We define 
an {\it affine perturbation} series about a function $g(x)$ as 
\beqan
\label{eq-affine-pert} 
f(x)&=&(A_0 +\cdots +A_p x^p+ \cdots)\, g(x)\\
&&\qquad\qquad+ \ (B_0 + \cdots + B_q x^q + \cdots), 
\eeqan
where $A_i$ and $B_i$ are constants with $p$ and $q$ positive 
rational numbers. Analogous to the Schwarzschild case, the
the bending angle has an invariant affine perturbation series of the form 
\beqa 
\label{affine-pert} 
\alphahat(b') & = & \Bigl(\sigma_0 +  \sigma_1 \, (b') + \sigma_2 \, (b')^2 
+ \cdots \, \Bigr) \, 
\log{\left[\frac{\lambda_0}{b'}\right]}\nonumber \\ 
& & \qquad \Bigl(\rho_0  + \rho_1 \, (b') 
+ \rho_2 \, (b')^2 
+ \cdots \, \Bigr), 
\eeqa 
where $\lambda_0$,  $\sigma_i$ and $\rho_i$ are not just numerical 
constants, but also depend on the spin parameter $a$.  Note 
that (\ref{affine-pert}) is not a Taylor series expansion because of the appearance of the 
logarithmic term.  

The formal expression (\ref{ExactKAlpha}) for the bending angle is just that; it is 
difficult to extract information about position and magnification of images as 
seen from the observer's vantage point.  The difference between the direct 
and retro orbits, for example, is simply not at all obvious if we just look at the 
formal exact expression.  The effect of frame-dragging on the light ray is buried deep
in the details of the expression.  In what follows, we have sorted through these
details being careful at each step to make sure that the Schwarzschild results are recovered 
whenever the spin parameter is turned off.  

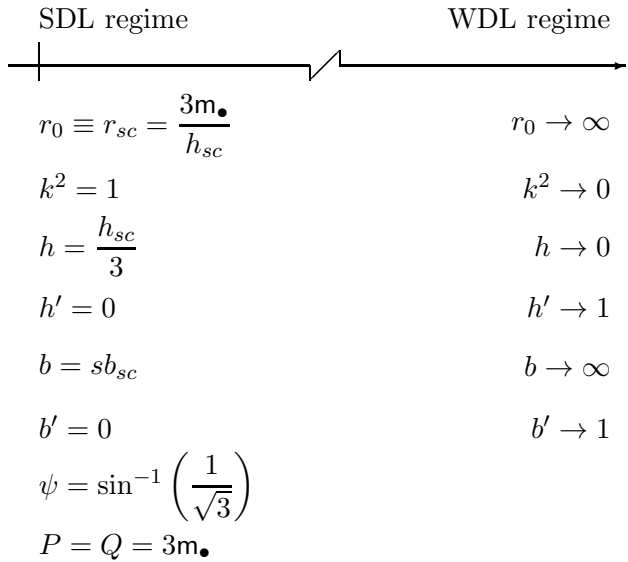
\begin{figure}[htp] 
\begin{center} 
\begin{center}
\setlength{\unitlength}{1mm}
\begin{picture}(80,80)(0,0)
\put(4,70){\makebox(0,0)[l]{SDL regime}}
\put(80,70){\makebox(0,0)[r]{WDL regime}}
\put(0,64){\line(1,0){40}}
\put(40,62){\line(0,1){2}}
\put(40,62){\line(1,1){4}}
\put(44,64){\line(0,1){2}}
\put(44,64){\vector(1,0){38}}
\put(4,62){\line(0,1){5}}
\put(4,56){\makebox(0,0)[l]{$r_0\equiv r_{sc}=\dfrac{3\mdot}{h_{sc}}$}}
\put(80,56){\makebox(0,0)[r]{$r_0 \rightarrow \infty$}}
\put(4,48){\makebox(0,0)[l]{$k^2=1$}}
\put(80,48){\makebox(0,0)[r]{$k^2 \rightarrow 0$}}
\put(4,40){\makebox(0,0)[l]{$h=\dfrac{h_{sc}}{3}$}}
\put(80,40){\makebox(0,0)[r]{$h \rightarrow 0$}}
\put(4,32){\makebox(0,0)[l]{$h'=0$}}
\put(80,32){\makebox(0,0)[r]{$h' \rightarrow 1$}}
\put(4,24){\makebox(0,0)[l]{$b=s b_{sc}$}}
\put(80,24){\makebox(0,0)[r]{$b \rightarrow \infty$}}
\put(4,16){\makebox(0,0)[l]{$b'=0$}}
\put(80,16){\makebox(0,0)[r]{$b' \rightarrow 1$}}
\put(4,8){\makebox(0,0)[l]{$\psi=\sin^{-1}\left(\dfrac{1}{\sqrt{3}}\right)$}}
\put(4,0){\makebox(0,0)[l]{$P=Q=3\mdot$}}
\end{picture}
\end{center}
\caption{\small \sl A schematic to show the strong and weak deflection limits in terms of the relevant variables.} 
\label{SDLandWDL} 
\end{center} 
\end{figure}

We begin by rewriting the expression in terms of the variables $h, h_{sc},\omega_s$, and $\omega_0$.  The two 
regimes that we are referring to as SDL and WDL involve different series expansions of 
elliptic integral of the third kind and these have to carried out using the built-in EllipticPi 
functions and their properties in Mathematica.  
The WDL is easily expressed as the limit as $b\rightarrow 0$, while a 
precise definition of the SDL is a bit more involved.  Following our scheme 
in \cite{iyerpetters}, we introduce the variable $b'=1-s b_{sc}/b$ and consider the 
limit as $b'\rightarrow 0$.  In this limit, we are approaching the critical impact parameter.  
Figure~\ref{SDLandWDL} is a schematic to show the limiting values of the different 
variables as the impact parameter goes from critical to infinity.  

On the left hand side, we have 
all parameters and the corresponding limiting values as  we approach critical 
radius.  First, $r_0$ is defined in a similar way as we did in the 
Schwarzschild case \cite{iyerpetters}.  The argument $k^2$ is defined through the elliptic
integral as usual in the range $0\le k^2 \le 1$.  Note that in some references the variable 
is referred to as $k^2$ and in others, simply as $k$.  The order in which the
arguments appear in $\Pi(n,\psi,k)$ also varies between different references 
and in Mathematica.  The variables $h'$ and $b'$ are both equivalent 
ways of describing the range of possible impact parameters.  We will eliminate $h'$ in the end to 
express all quantities in terms of $b'$ in the end.  The impact parameter $b$ itself can be positive 
or negative and the sign is carried by $s$.  Note that the impact parameter explicitly approaches 
critical on the left and infinity on the right.  Some of the other intermediate variables are also given
here for completeness.  Once again, our goal is to express the bending angle in terms of $b'$; the only other
quantities that will remain in our final results are $\mdot$, $\hat{a}$ and $s$.  We have assumed that
 $\hat{a}$ stays in the range $0\le\hat{a}<1$.  In addition to the variables shown in 
Figure \ref{SDLandWDL}, the argument $n_\pm$ also stays in the range $\{0\rightarrow 1\}$.  

\section{Weak Deflection Limit}

The weak deflection limit is simply defined by the limit as $b\rightarrow \infty$.  Series expansions
for the elliptic integrals are readily available in Mathematica.  Before we present our result, we give here
the existing correction to the Einstein bending angle that other authors have derived before.  For example,
the following result appeared in 1967 in one of the earliest papers on the Kerr metric by
Boyer and Lindquist \cite{boy-lind}:
\beq
\label{boy-lind}
\alphahat=4\frac{\mdot}{r_0}\left(1+\frac{a}{r_0}\right)
\eeq
The $r_0$ is the Boyer-Lindquist ``radial" coordinate or ``the distance of closest approach".  This 
result is for direct orbits only and it clearly shows that there is no spin-dependent correction to the Einstein angle 
to first order in $1/b$.  In the above result and in a number of other references, the direct orbit is one in which
$a$ is taken to be negative; this is a matter of convention.  Similar results (some with the opposite sign convention) were
also found by Skrotskii\cite{skrotskii} and Plebanski\cite{plebanski} and others.  Extensive numerical analysis can
be found in Rauch and Blandford \cite{rauch-blandford}.

As in the Schwarzschild case, we have chosen to keep the expression in terms of the invariant
quantity, $b$ instead of $r_0$.  After a straightforward calculation and
some simplification we obtain the following bending angle WDL series for the equatorial Kerr:
\begin{widetext}
\beqa
\label{AlphaKWeak} 
\alphahat=4\left(\frac{\mdot}{b}\right)+\left[\frac{15\pi}{4}
-4s\hat{a}\right] \left(\frac{\mdot}{b}\right)^2
&+&\left[\frac{128}{3}-10\pi s\hat{a}+4\hat{a}^2\right] \left(\frac{\mdot}{b}\right)^3 \nonumber\\
&&\qquad+\left[\frac{3465\pi}{64}-192s\hat{a}+\frac{285\pi}{16}\hat{a}^2-4s\hat{a}^3\right]
\left(\frac{\mdot}{b}\right)^4+...
\eeqa
\end{widetext}
The spin parameter $\hat{a}$ and the sign $s$ appear explicitly in our series expansion.  
We will first note that the leading behavior of our result is in agreement with (\ref{boy-lind}) after 
correcting for the sign; specifically, we do not see a correction to first order in $1/b$.  
As mentioned in the introduction, the leading terms of our series are in exact agreement with those obtained 
by Petters in a completely different context \cite{petters}.  Note that some of the terms that 
depend on the spin parameter are also sign sensitive.  Although it may never be needed,
the weak deflection series can be extended to include terms of much higher orders, for 
different values of $\hat{a}$ and for both direct and retro orbits.  As a first check of our result,
we set $a=0$ in equation (\ref{AlphaKWeak}) 
\begin{widetext}
\beq
\alphahat (b)=4\left(\frac{\mdot}{b}\right) +  
\frac{15\pi}{4}\left(\frac{\mdot}{b}\right)^2
+\frac{128}{3}\left(\frac{\mdot}{b}\right)^3+\frac{3465\pi}{64}\left(\frac{\mdot}{b}\right)^4+
\mathcal{O}\left[\left(\frac{\mdot}{b}\right)^5\right]+...
\eeq
\end{widetext}
and recover the Schwarzschild series \cite{iyerpetters}.

\begin{figure}[htp] 
\begin{center} 
\includegraphics[width=3.3in]{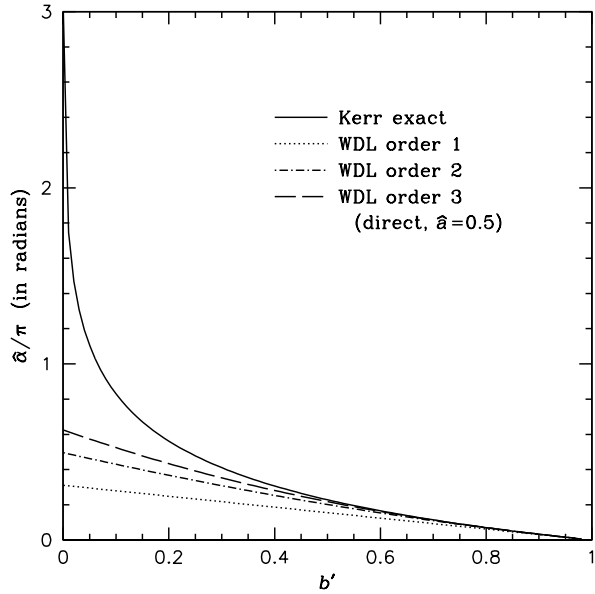}
\caption{\small \sl Plot comparison of perturbative and exact angle for $\hat{a}=0.5$ direct orbit.} 
\label{WDLD123p5} 
\end{center} 
\end{figure}

\begin{figure}[htp] 
\begin{center} 
\includegraphics[width=3.3in]{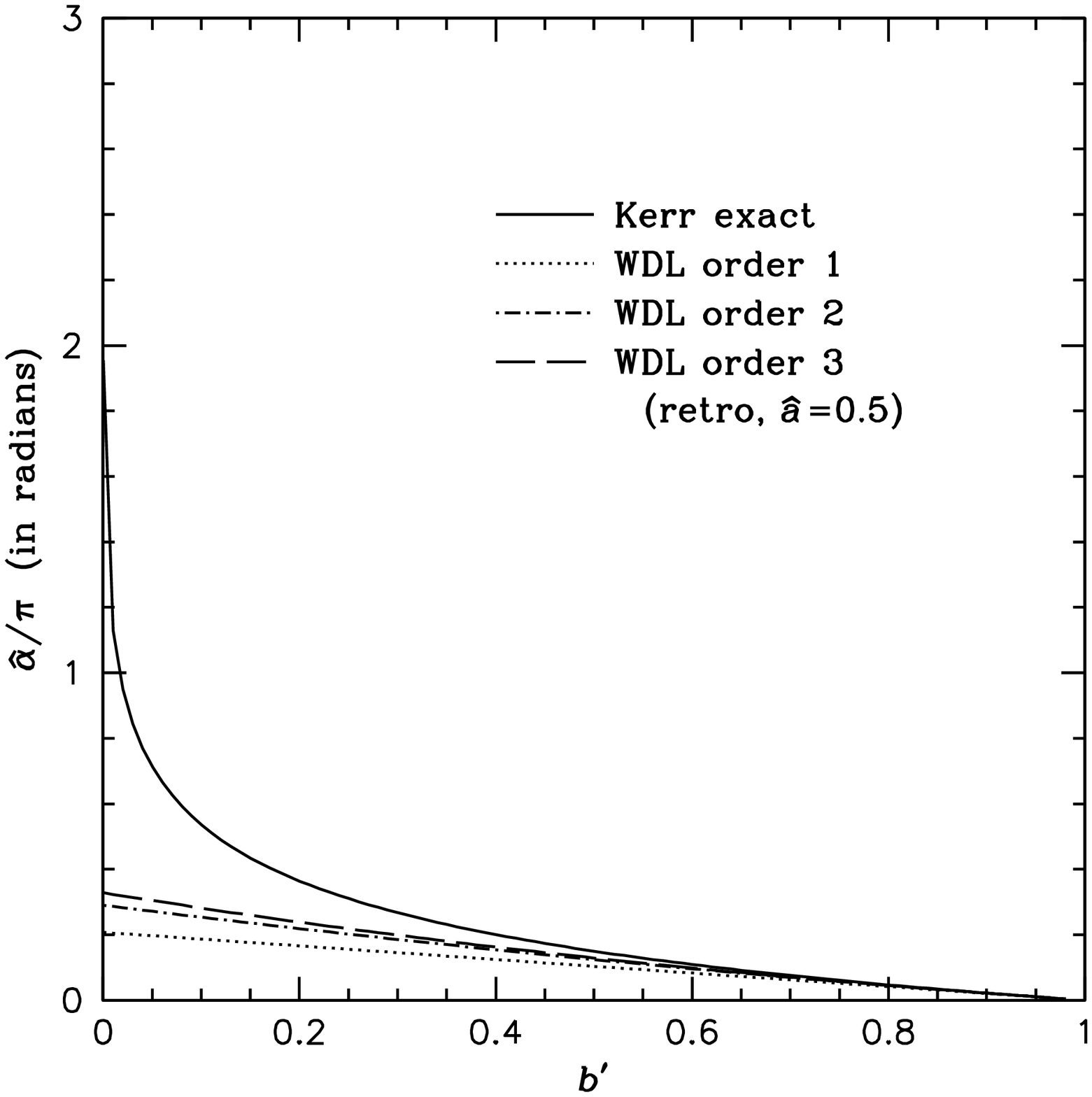}
\caption{\small \sl Plot comparison of perturbative and exact angle for $\hat{a}=0.5$ retro orbit.} 
\label{WDLR123p5} 
\end{center} 
\end{figure}

\begin{figure}[htp] 
\begin{center} 
\includegraphics[width=3.3in]{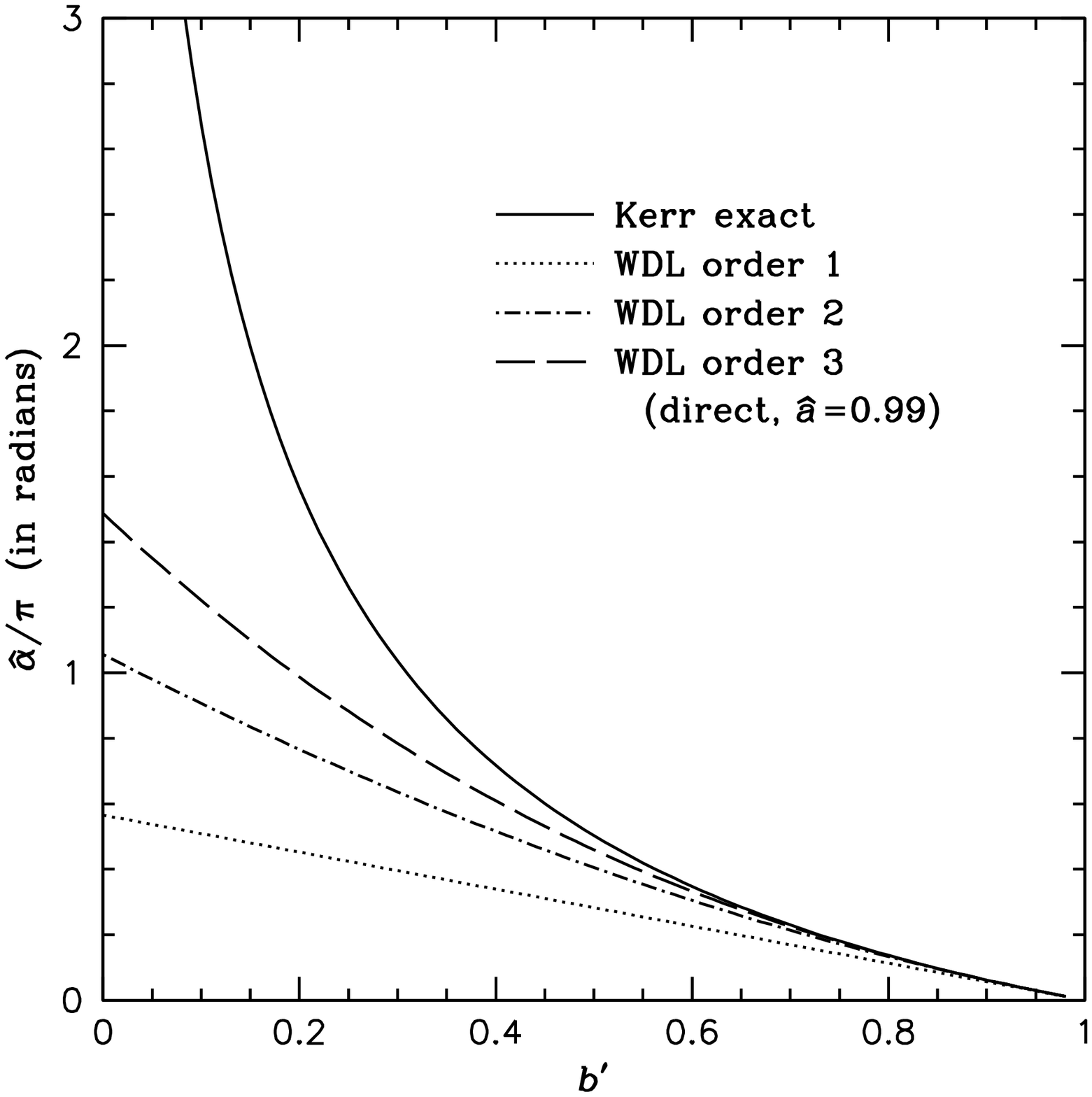}
\caption{\small \sl Plot comparison of perturbative and exact angle for $\hat{a}=0.99$ direct orbit.} 
\label{WDLD123p99} 
\end{center} 
\end{figure}

\begin{figure}[htp] 
\begin{center} 
\includegraphics[width=3.3in]{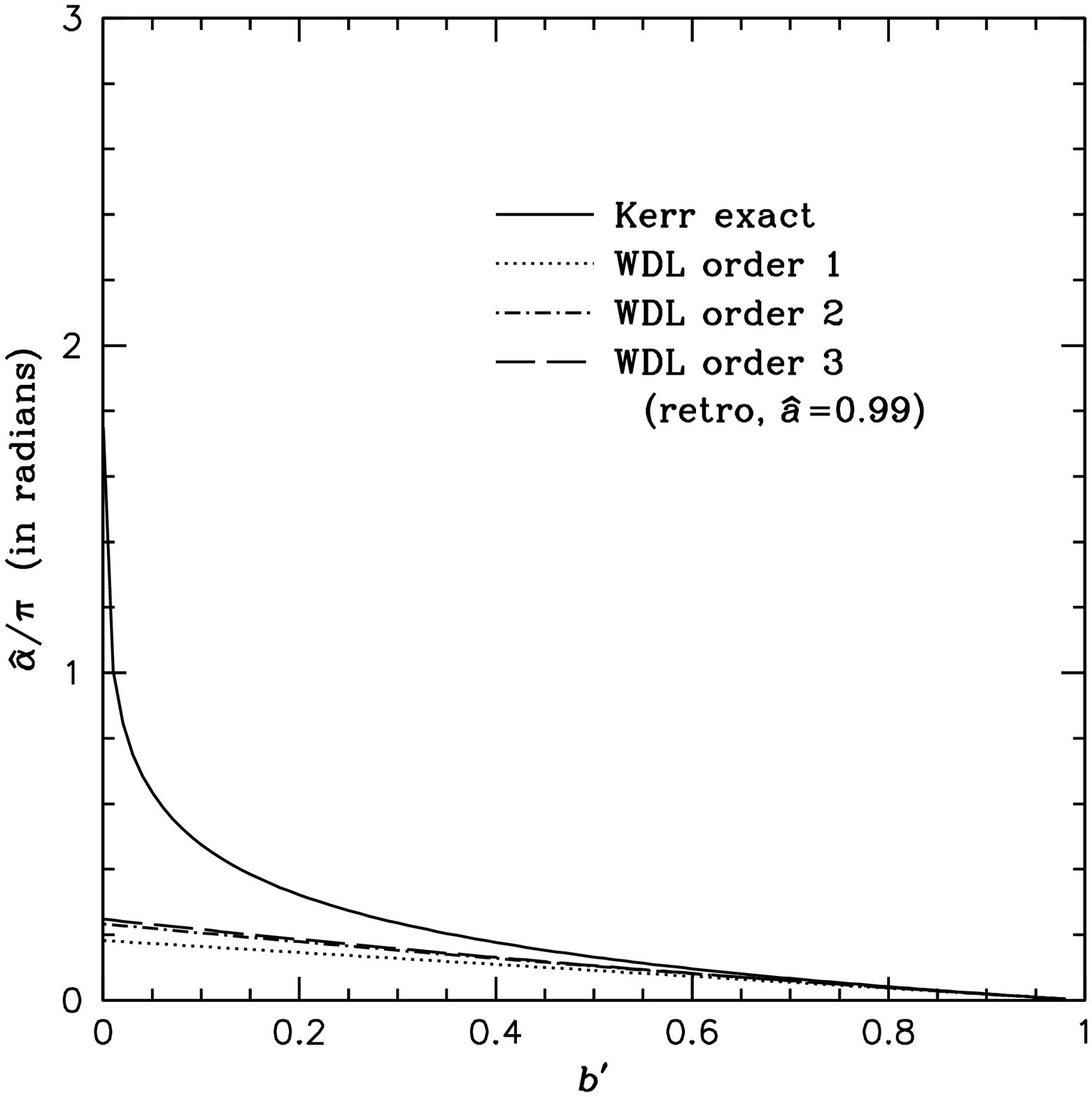}
\caption{\small \sl Plot comparison of perturbative and exact angle for $\hat{a}=0.99$ retro orbit.} 
\label{WDLR123p99} 
\end{center} 
\end{figure}

In Figures \ref{WDLD123p5}-\ref{WDLR123p99}
plot comparisons of the perturbative and exact Kerr bending angle are presented for $\hat{a}=0.5$ and $\hat{a}=0.99$ 
for both direct and retro orbits.  Corrections up to first, second and third order in $\mdot/b$ are plotted with 
the exact bending angle in each case.  The plots clearly show that the weak deflection approximation 
gets closer to the exact angle as we include higher order terms.  We have shown direct and retro plots to
illustrate the effectiveness of the weak deflection series for both $s=+1$ and $s=-1$.  We have chosen two 
representative values of $\hat{a}$ for the plots; in fact, the series works nicely for all spins $0\le\hat{a}<1$.

The correction to the Schwarzschild weak deflection bending angle coming from spin 
appears in second order ($1/b^2$) only.  This means that the spin contribution
would be nearly impossible to detect in the weak deflection regime.  The Einstein ring, for
example, when the source, lens and observer are perfectly aligned would be essentially the same for
the Schwarzschild and Kerr case, making it extremely difficult to detect the spin contribution.  As we will see shortly,
the formation of relativistic images in the strong deflection limit is much more sensitive to the spin and for this 
reason would be a far more important tool for testing general relativity in the strong field 
limit and for studying the black hole spin itself.

\section{Strong Deflection Limit}

As we approach critical impact parameter, we once again see, as in the 
Schwarzschild case \cite{iyerpetters}, that the leading behavior is logarithmic.  In addition
to this virulent term, the dependence on spin makes the series expansion quite complicated.  We have
nevertheless been able to extract the series terms up to second order in $b'$ as defined in 
equation (\ref{affine-pert}).  The built-in series expansions in Mathematica have been used 
for this purpose\cite{mathematica}.  Care must be taken with defining and providing the correct arguments for the 
Mathematica functions and appropriate limits for the associated series.  We 
give here only the zeroth order term for the bending angle:

\begin{widetext}
\beq
\label{zeroth}
\alphahat_{\rm SDL}^{(0)}=-\pi+\frac{3\sqrt{\dfrac{1}{h_{sc}}}\left[2\sqrt{1-\omega_0}
\left[3-2h_{sc}(1-\omega_s)\right] \log \left(\dfrac{12(2-\sqrt{3})}{h'}\right)+\sqrt{3}
\Bigl[U_-V_-+U_+V_+\Bigr]\right]}{\sqrt{1-\omega_0}\left[9-h_{sc}(6-h_{sc}\omega_0)\right](1-\omega_s)}\\\nonumber
\eeq
where
\beqan
U_{\pm}\equiv &&h_{sc}\left[\pm\omega_0\mp2(1-\omega_s)\left(1\pm\sqrt{1-\omega_0}
\right)\right]\pm3\left[1\pm\sqrt{1-\omega_0}-2\omega_s\right]\\
\\
V_{\pm}=&&\sqrt{\frac{h_{sc}\omega_0}{6\pm6\sqrt{1-\omega_0}+h_{sc}\omega_0}} 
\log\left[\frac{\left(1+\sqrt{\dfrac{h_{sc}\omega_0}{6\pm6\sqrt{1-\omega_0}+h_{sc}\omega_0}} 
\right)\left(1-\sqrt{\dfrac{3h_{sc}\omega_0}{6\pm6\sqrt{1-\omega_0}+h_{sc}\omega_0}} \right)}
{\left(1-\sqrt{\dfrac{h_{sc}\omega_0}{6\pm6\sqrt{1-\omega_0}+h_{sc}\omega_0}} \right)
\left(1+\sqrt{\dfrac{3h_{sc}\omega_0}{6\pm6\sqrt{1-\omega_0}+h_{sc}\omega_0}} \right)}\right]\\
\\
\eeqan
\end{widetext}

We have written the bending angle in terms of $h'$ because the full version explicitly in terms of $b'$ is too 
complicated to show here.  It can be shown, almost just by inspection of the above zeroth order term, that setting 
$\hat{a}$ equal to zero (which means $\omega_s,\omega_0\rightarrow 0$ and $h_{sc}\rightarrow 1$)
yields the Schwarzschild version \cite{iyerpetters} immediately:
\beq 
\alphahat
= -\pi + 2 \log\left(\frac{12(2-\sqrt{3})}{h'}\right) + \mathcal{O}[h']+...
\eeq 

\begin{figure}[htp] 
\begin{center} 
\includegraphics[width=3.3in]{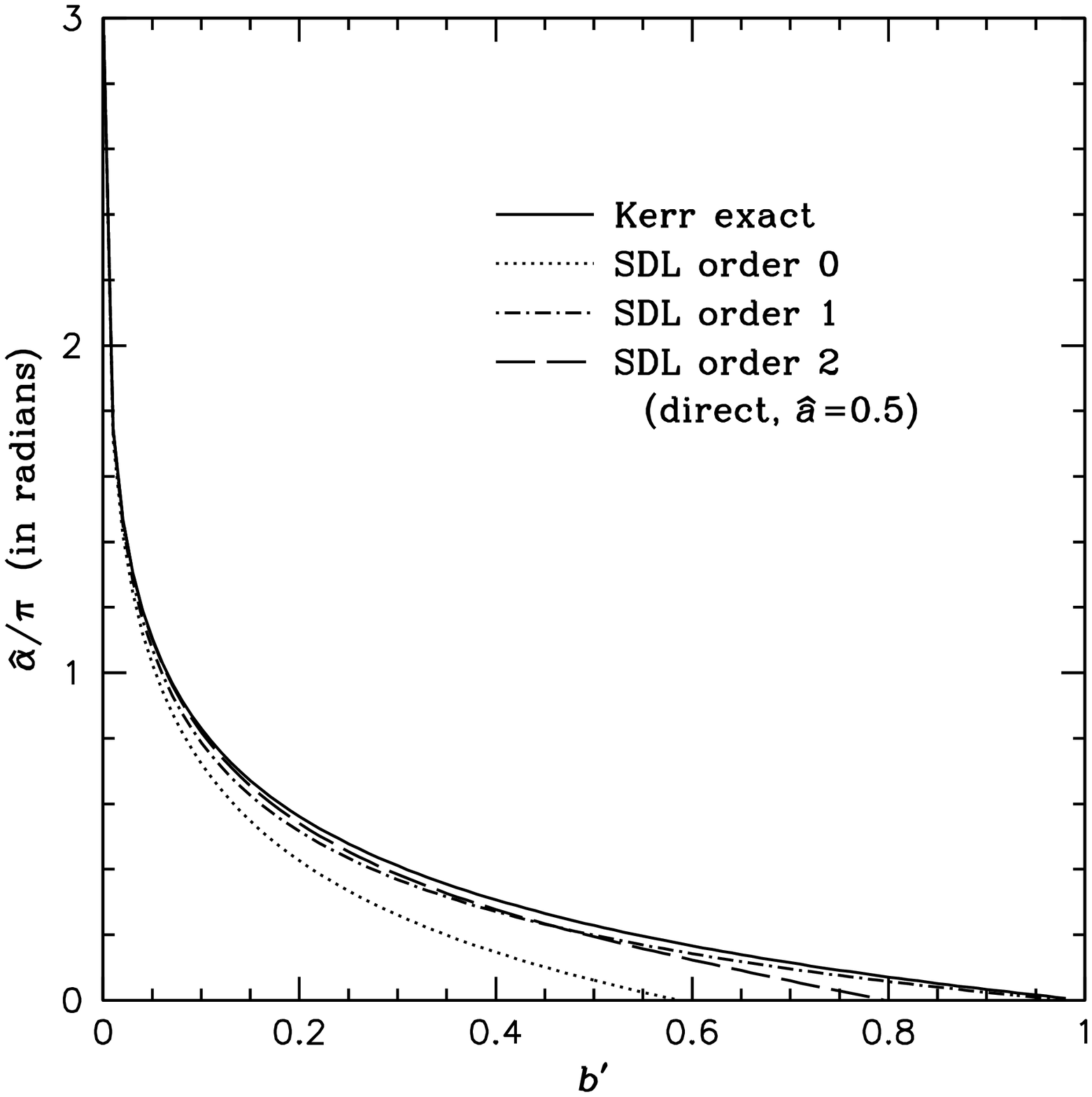}
\caption{\small \sl Plot comparison of perturbative and exact angle for $\hat{a}=0.5$ direct orbit.} 
\label{SDLD012p5} 
\end{center} 
\end{figure}

\begin{figure}[htp] 
\begin{center} 
\includegraphics[width=3.3in]{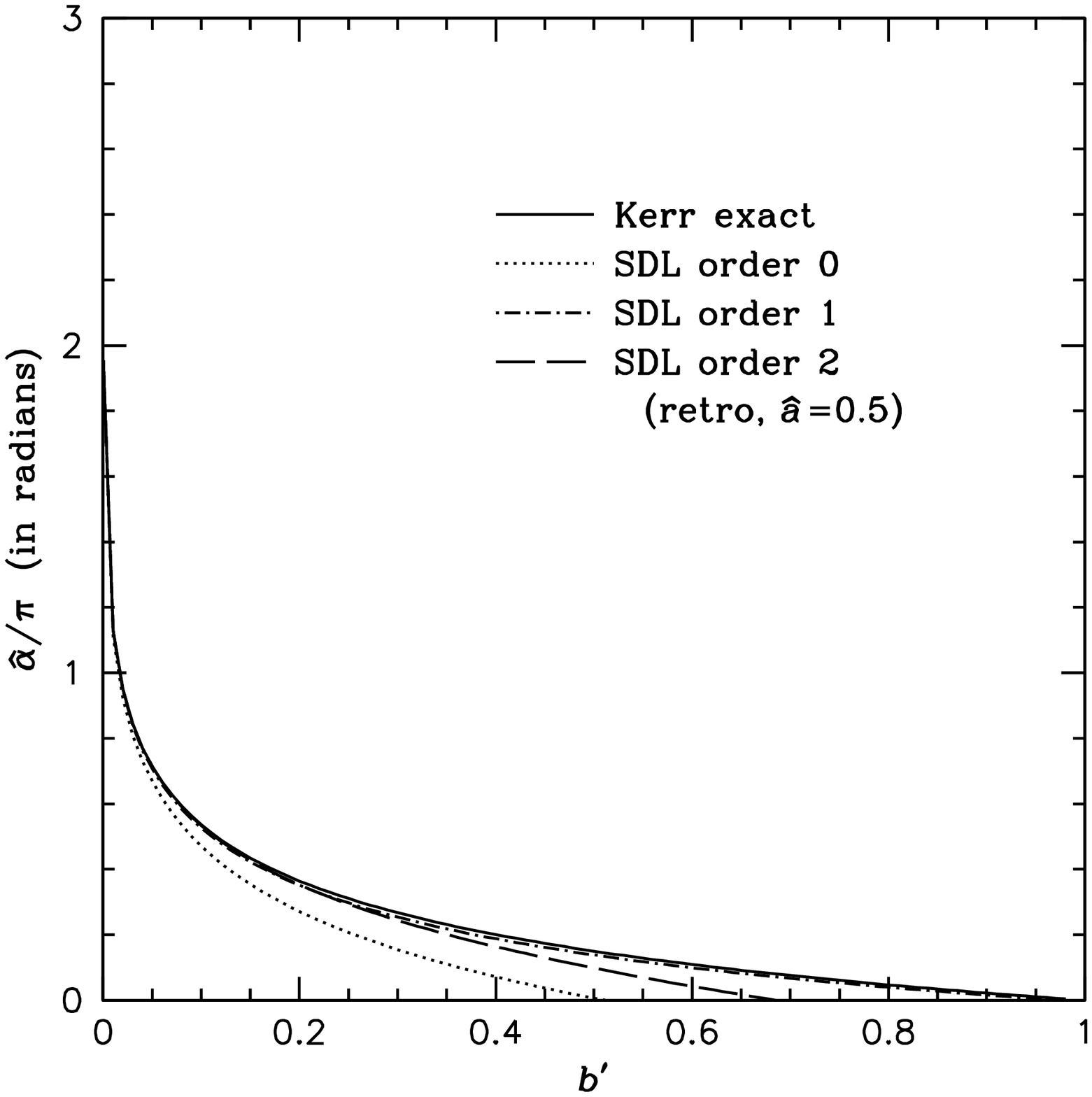}
\caption{\small \sl Plot comparison of perturbative and exact angle for $\hat{a}=0.5$ retro orbit.} 
\label{SDLR012p5} 
\end{center} 
\end{figure}

\begin{figure}[htp] 
\begin{center} 
\includegraphics[width=3.3in]{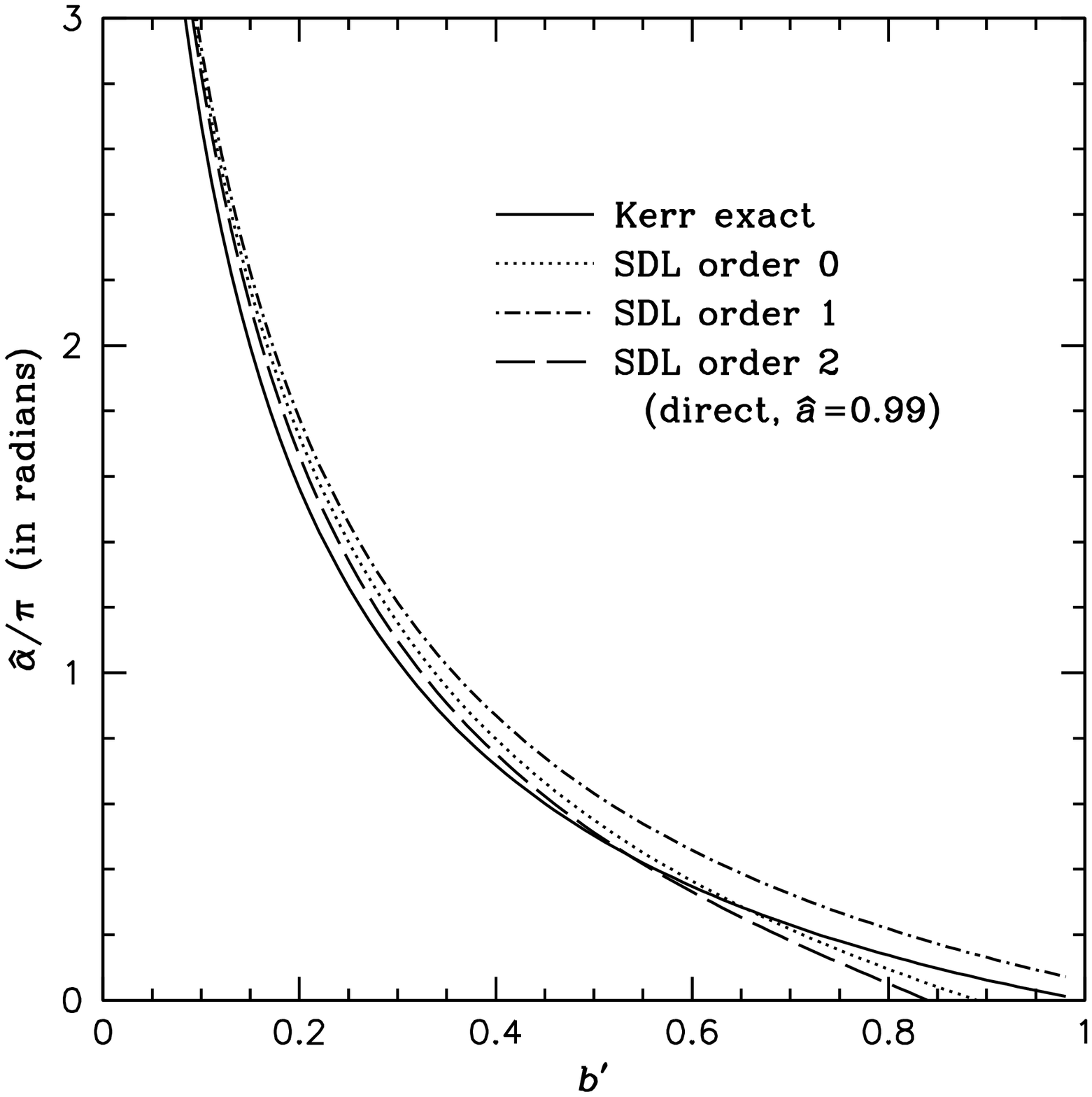}
\caption{\small \sl Plot comparison of perturbative and exact angle for $\hat{a}=0.99$ direct orbit.} 
\label{SDLD012p99} 
\end{center} 
\end{figure}

\begin{figure}[htp] 
\begin{center} 
\includegraphics[width=3.3in]{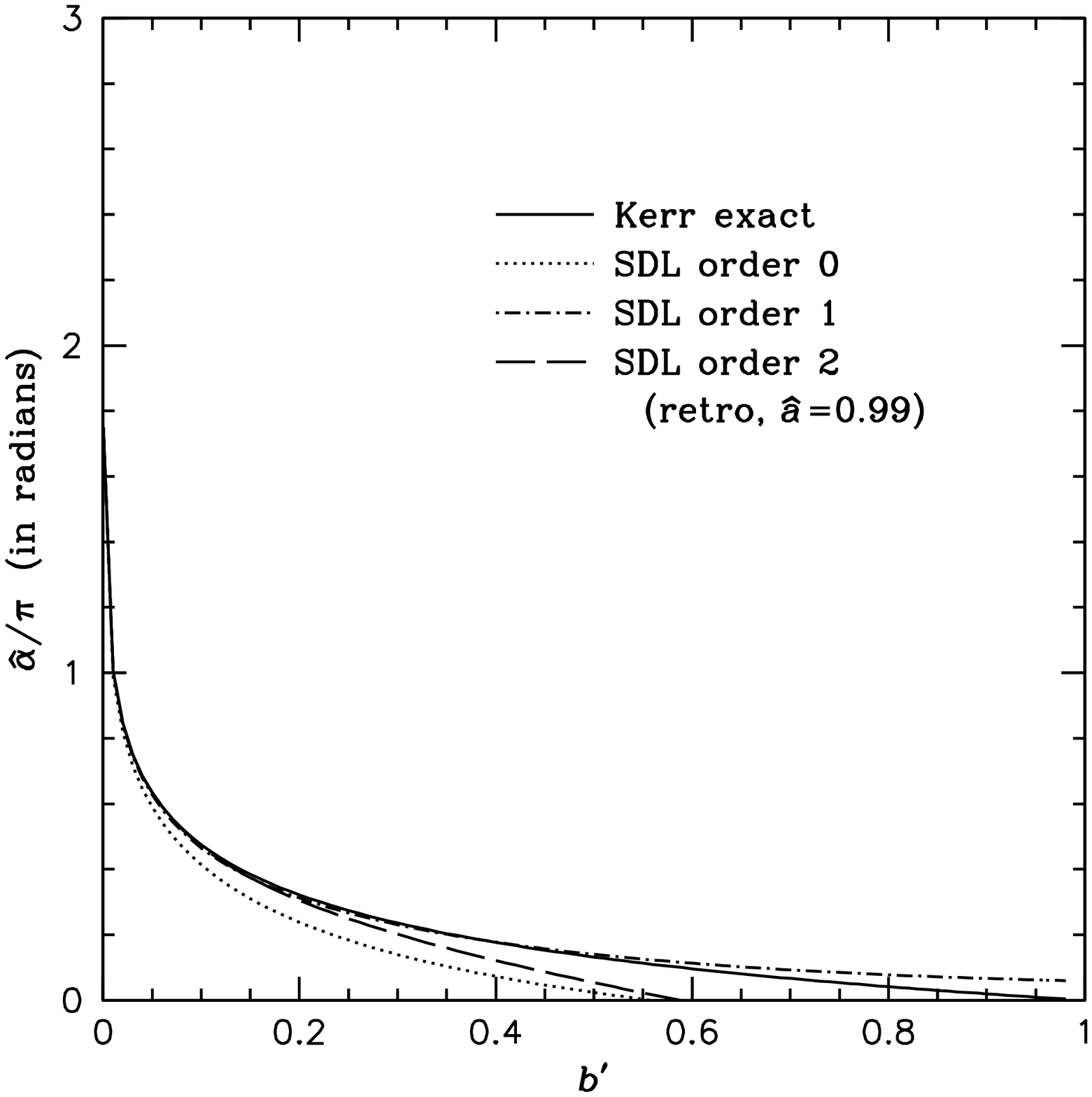}
\caption{\small \sl Plot comparison of perturbative and exact angle for $\hat{a}=0.99$ retro orbit.} 
\label{SDLR012p99} 
\end{center} 
\end{figure}

In Figures \ref{SDLD012p5}-\ref{SDLR012p99} are shown plots
of the strong deflection approximation and the exact bending angle.  Due to the presence of both
positive and negative terms in the series, the perturbative expression 
oscillates about the exact value and gets closer to the exact as we include more terms in the 
expansion.

\section{Einstein Ring and Para Images}

\begin{figure}[htp] 
\begin{center} 
\includegraphics[width=3.3in]{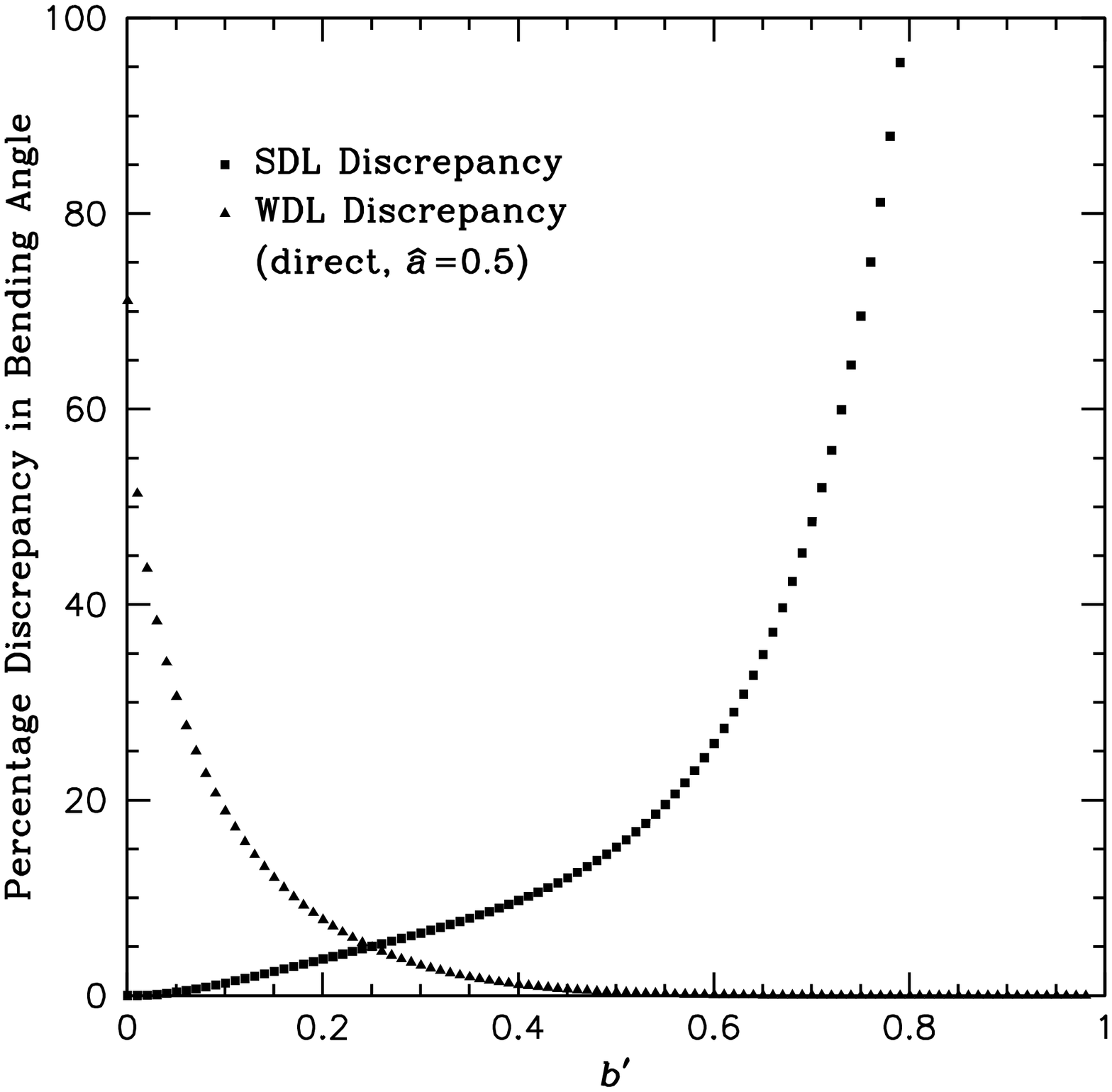}
\caption{\small \sl Percentage discrepancy for 2nd-order strong deflection and the 6th-order 
weak deflection for direct orbit with $\hat{a}=0.5$.  The horizontal axis corresponds to the exact value.} 
\label{discreDp5} 
\end{center} 
\end{figure}

\begin{figure}[htp] 
\begin{center} 
\includegraphics[width=3.3in]{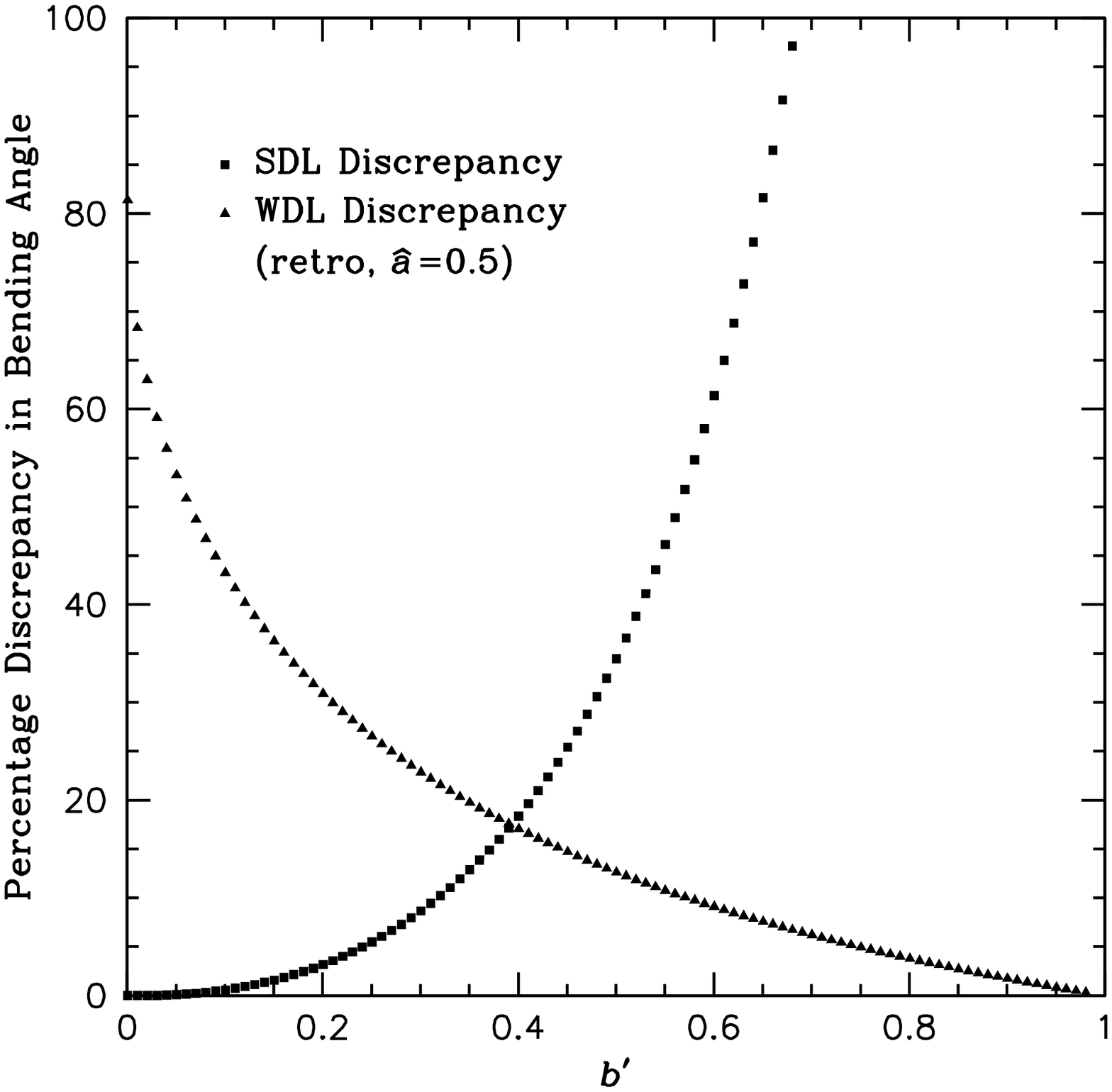}
\caption{\small \sl Percentage discrepancy for 2nd-order strong deflection and the 6th-order 
weak deflection for retro orbit with $\hat{a}=0.5$.  The horizontal axis corresponds to the exact value.} 
\label{discreRp5} 
\end{center} 
\end{figure}

The weak and strong series expansions are applicable to either ends of the 
entire range of $b'$.  When both the WDL and SDL series are taken together, they cover the entire range
with accuracy.  Figures \ref{discreDp5} and \ref{discreRp5} depict this for the case when $\hat{a}=0.5$.  The value of $b'$
where the discrepancy plots criss-cross varies, depending on the spin 
value.  In Figures \ref{BestDp5} and \ref{BestRp5}, we showed plots of
the 2nd-order strong series and the 6th-order weak series along with the exact for spin value of $\hat{a}=0.5$.  From
all these plots it is clearly evident that our series expansion results are excellent approximations to the exact 
bending angle.  The is essentially the same for all spin values between $0\le\hat{a}<1$.  

We present here an application of the strong deflection limit of the bending angle used
along with the lens equation for the lens geometry shown in Figure~\ref{LensGeom} with the usual assumption
that the thickness of the lens plane is negligible compared to the distances between the source, the 
lens and the observer.  For the calculations shown here we will use the same model
as in \cite{vir-ellis}, where the mass of black hole is taken to be $2.8\times10^6\,M_{\odot}$.  With
$D_{LS}/D_S=1/2$ and $D_L=8.5\;{\rm kpc}$, the angular radius of the Einstein ring can be calculated
using $\theta_E=\left(4\mdot D_{LS}/D_LD_S\right)^{1/2}$.  The Einstein ring is formed by a cone's worth of rays emanating from the 
source and symmetrically deflected by the black hole.  In the Kerr case, we expect that the projection of these rays 
on the sky would still be a closed curve, except it won't have circular symmetry; depending on how the magnification
varies, these could appear as arcs on the sky.  Furthermore, our results 
apply only to light rays that stay on the equatorial plane, and hence the plane of 
incidence.  Rays coming in at an angle to the equatorial plane, however, will not stay 
on the plane of incidence: the black hole spin will peel these away from the plane of
incidence (see \cite{vazquez-esteban}and \cite{bozza}).   Since we are working with
just the equatorial Kerr case, our results pertain to only two of these rays, one on each
side.  If these were the only two rays emanating from the source, we would see two point 
images on either side of the black hole.  These are basically the antipodal points of the ``ring" image that
would be formed.  We refer to these image positions as ``para images"--- ``stardogs" or ``parastars" are terms that 
would be appropriate as well.  Angular positions for the para images are listed for the case when the 
source, lens and observer are perfectly aligned (i.e., $\beta=0$). Table \ref{stardogs} is analogous to Table
III in \cite{vir-ellis}.  (The small discrepancies in the $\hat{a}=0$ data compared to those obtained by
\cite{vir-ellis} is due to rounding off in the physical constant and in conversion factors.) 
We use the weak deflection limit expansion for calculating $\theta_E$ for the 
stardog positions on the sky.  The angular separation between the para, or stardog positions in the Kerr case
is analogous to diameter of the Einstein ring in Schwarzschild geometry.  

\begin{figure}[htp] 
\begin{center} 
\includegraphics{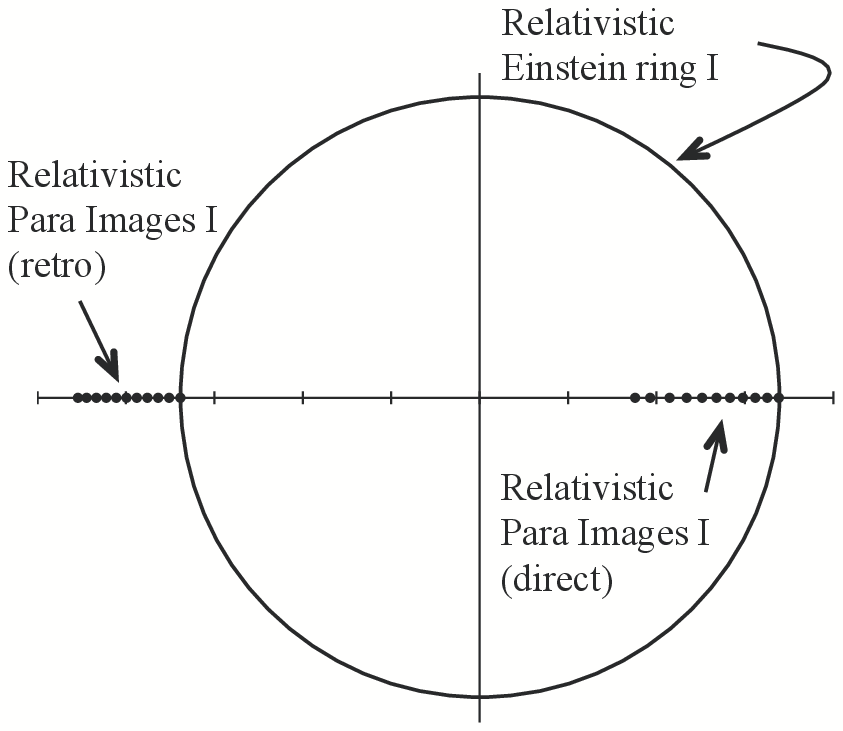}
\caption{\small \sl Para positions or antipodal points on either side of the Kerr black hole where 
relativistic image I would appear on the sky, are shown as a function of increasing $\hat{a}$.  The data 
used for this figure can be found in Table \ref{stardogs}.} 
\label{paraimagesI} 
\end{center} 
\end{figure} 

\begin{figure}[htp] 
\begin{center} 
\includegraphics{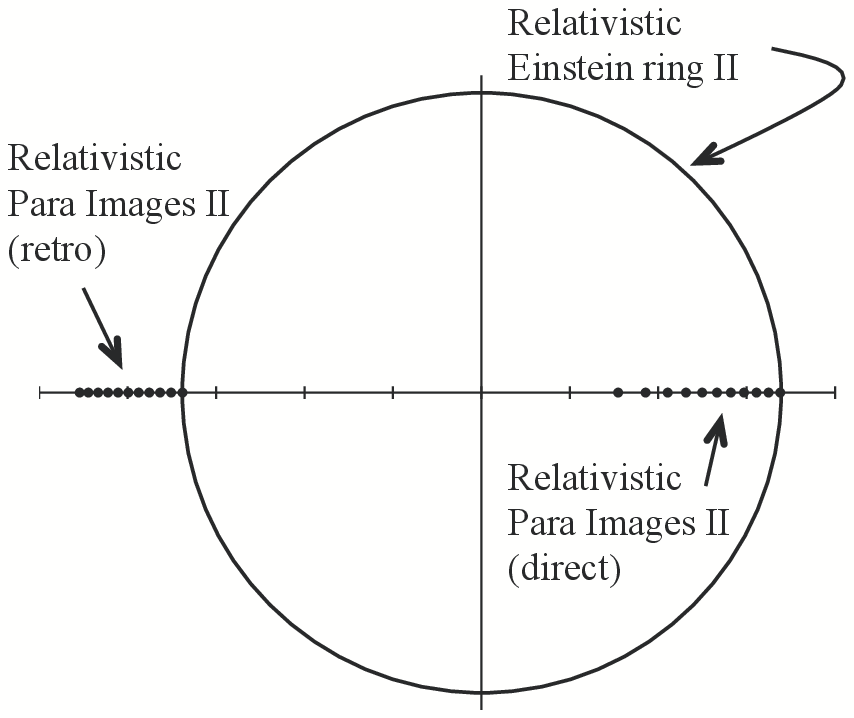}
\caption{\small \sl Para positions or antipodal points on either side of the Kerr black hole where 
relativistic image I would appear on the sky, are shown as a function of increasing $\hat{a}$. The data 
used for this figure can be found in Table \ref{stardogs}.} 
\label{paraimagesII} 
\end{center} 
\end{figure}

\begin{table*}
\scalebox{0.6}{\begin{tabular}{lccccc}
\hline
\hline
Rings/Para images&$a$&$\theta$&$\hat{\alpha}$&${r_0}/{2\mdot}$&${r_0}/{r_{\rm crit}}$\\
\hline
Einstein ring&0&1.1583 arcsec&2.3166 arcsec&178074& \\
Para image:& $\quad$(all spins)$\quad$ &$\quad$1.1583 arcsec$\quad$&$\quad2.3166$ arcsec$\quad$&$\quad178074\quad$& \\
\hline
Relativistic Einstein ring I&0&$16.9207\;\mu$as&$2\pi+33.8415\;\mu$as&$1.54505$&$\quad1.03004\quad$\\
Relativistic Para image I:&$0.000001$&$16.9203\;\mu$as&$2\pi+33.8407\;\mu$as&1.54462&1.02974\\
$\qquad$(direct)&0.1&$16.2649\;\mu$as&$2\pi+32.5299\;\mu$as&1.48908&1.03330\\
&0.2&$15.5890\;\mu$as&$2\pi+31.1780\;\mu$as&1.43143&1.03757\\
&0.3&$14.8888\;\mu$as&$2\pi+29.7776\;\mu$as&1.37131&1.04281\\
&0.4&$14.1595\;\mu$as&$2\pi+28.3191\;\mu$as&1.30821&1.04936\\
&0.5&$13.3943\;\mu$as&$2\pi+26.7885\;\mu$as&1.24142&1.05774\\
&0.6&$12.5829\;\mu$as&$2\pi+25.1659\;\mu$as&1.16982&1.06886\\
&0.7&$11.7094\;\mu$as&$2\pi+23.4187\;\mu$as&1.09156&1.08433\\
&0.8&$10.7447\;\mu$as&$2\pi+21.4894\;\mu$as&1.00335&1.10801\\
&0.9&$9.6445\;\mu$as&$2\pi+19.2891\;\mu$as&0.90193&1.15792\\
&0.99&$8.7941\;\mu$as&$2\pi+17.5881\;\mu$as&0.84449&1.44648\\
Relativistic Para image I:&0.000001&$-16.9204\;\mu$as&$2\pi+33.8407\;\mu$as&1.54462&1.02974\\
$\qquad$(retro)&0.1&$-17.5581\;\mu$as&$2\pi+32.1162\;\mu$as&1.59832&1.02675\\
&0.2&$-18.1804\;\mu$as&$2\pi+36.3608\;\mu$as&1.65041&1.02420\\
&0.3&$-18.7891\;\mu$as&$2\pi+37.5783\;\mu$as&1.70106&1.02201\\
&0.4&$-19.3858\;\mu$as&$2\pi+38.7716\;\mu$as&1.75042&1.02011\\
&0.5&$-19.9716\;\mu$as&$2\pi+39.9432\;\mu$as&1.79861&1.01844\\
&0.6&$-20.5477\;\mu$as&$2\pi+41.0953\;\mu$as&1.84573&1.01697\\
&0.7&$-21.1148\;\mu$as&$2\pi+42.2296\;\mu$as&1.89187&1.01567\\
&0.8&$-21.6739\;\mu$as&$2\pi+43.3477\;\mu$as&1.93710&1.01452\\
&0.9&$-22.2255\;\mu$as&$2\pi+44.4510\;\mu$as&1.98149&1.01348\\
&0.99&$-22.7161\;\mu$as&$2\pi+45.4321\;\mu$as&2.02077&1.01264\\
\hline
Relativistic Einstein ring II&0&$16.8996\;\mu$as&$4\pi+33.7992\;\mu$as&$1.50188$&$\quad1.00125\quad$\\
Relativistic Para image  II:&$0.000001$&$16.8996\;\mu$as&$4\pi+33.7992\mu$as&1.50187&1.00125\\
$\qquad$(direct)&0.1&$16.2393\;\mu$as&$4\pi+32.4785\mu$as&1.44337&1.00158\\
&0.2&$15.5568\;\mu$as&$4\pi+31.1135\;\mu$as&1.38239&1.00203\\
&0.3&$14.8477\;\mu$as&$4\pi+29.6954\;\mu$as&1.31850&1.00265\\
&0.4&$14.1060\;\mu$as&$4\pi+28.2120\;\mu$as&1.25112&1.00356\\
&0.5&$13.3230\;\mu$as&$4\pi+26.6460\;\mu$as&1.17944&1.00494\\
&0.6&$12.4853\;\mu$as&$4\pi+24.9707\;\mu$as&1.10226&1.00713\\
&0.7&$11.5708\;\mu$as&$4\pi+23.1417\;\mu$as&1.01765&1.01091\\
&0.8&$10.5376\;\mu$as&$4\pi+21.0752\;\mu$as&0.92208&1.01826\\
&0.9&$9.2856\;\mu$as&$4\pi+18.5713\;\mu$as&0.80701&1.03606\\
&0.99&$7.7345\;\mu$as&$4\pi+15.4690\;\mu$as&0.67348&1.15357\\
Relativistic Para image II:&$0.000001$&$-16.8996\;\mu$as&$4\pi+35.7992\mu$as&1.50188&1.00125\\
$\qquad$(retro)&0.1&$-17.5411\;\mu$as&$4\pi+35.0822\;\mu$as&1.55824&1.00100\\
&0.2&$-18.1664\;\mu$as&$4\pi+36.3328\;\mu$as&1.61272&1.00082\\
&0.3&$-18.7774\;\mu$as&$4\pi+37.5549\;\mu$as&1.66554&1.00067\\
&0.4&$-19.3760\;\mu$as&$4\pi+38.7519\;\mu$as&1.71688&1.00056\\
&0.5&$-19.9633\;\mu$as&$4\pi+39.9265\;\mu$as&1.76687&1.00047\\
&0.6&$-20.5405\;\mu$as&$4\pi+41.0811\;\mu$as&1.81564&1.00039\\
&0.7&$-21.1087\;\mu$as&$4\pi+42.2174\;\mu$as&1.86329&1.00033\\
&0.8&$-21.6686\;\mu$as&$4\pi+43.3372\;\mu$as&1.90992&1.00028\\
&0.9&$-22.2209\;\mu$as&$4\pi+44.4418\;\mu$as&1.95561&1.00024\\
&0.99&$-22.7120\;\mu$as&$4\pi+45.4240\;\mu$as&1.99598&1.00021\\
\hline
\hline
\end{tabular}}
\caption{\small \sl Para positions of Einstein and relativistic Einstein rings and the analogous images in Kerr geometry.  Images
on the retro side are indicated with negative values for the $\theta$ in microarcseconds ($\mu as$).} 
\label{stardogs}
\end{table*}

In the strong deflection limit, i.e., when the impact parameter 
is close to critical, we see multiple loops of the light ray as can be seen from the plot of 
the exact bending angle.  Rays that loop around multiple times result in multiple relativistic para images on 
either side just as in the Schwarzschild case.    The image positions we obtain
for direct orbits seem reasonable when compared with the higher order images studied by others
for the non-zero spins for direct orbits off the equatorial plane (see Table I of \cite{vazquez-esteban} 
and \cite{bozza} and references within.)  We believe that image predictions for retro orbits 
for all spin values $0\le \hat{a}<1$ have not appeared in literature before.  Image positions for direct versus 
retro orbits illustrates the key difference in the effect the black hole spin has on rays that 
are traversing upstream versus downstream: the para images are pushed outward 
on the retro side and inward on the prograde 
side.  The data used to generate Figures \ref{paraimagesI} 
and \ref{paraimagesII} are shown in Table \ref{stardogs}.  Positive and negative values of 
the angular position $\theta$ of the image indicate respectively whether the 
image is on the right or left side.   The last column in Table \ref{stardogs} gives the ratio 
$r_0/r_{\rm crit}$, rather than $r_0/2\mdot$ showing clearly that the rays are 
just outside critical radii.  We would like to reiterate here the 
importance of analytical expansion for the bending angle, without which
calculation of these image positions would not be possible.

\section{Conclusions}

Analytic solutions in the form of series expansions in the strong and weak deflection
regimes for the bending angle of light rays on the equatorial plane of the Kerr
black hole were presented in this paper.  Higher order terms in strong deflection limit
were not shown because they are too complicated.  The technique to generate these, however,
have been presented in detail.  We have also shown that both expansions have a high 
level of accuracy compared to the exact bending angle, and so would be excellent 
predictors of image positions.  To illustrate this, both expansions were applied
to the simplest gravitational lensing situation when the source, lens and observer are perfectly aligned 
on the optic axis.  We showed that the effect of the black hole spin distorts and 
shifts the Einstein ring inward on the direct side and outward
on the retrograde side.  For large spins, the shift in the image positions is much higher
than the angular separation of successive relativistic images.  Of course, whether this shift 
can be observed with precision will depend on the resolution power of future telescopes.  The series
expansions will be applied next to the case when the source is not on the optic axis and also calculations
pertaining to magnification of the para images.  In order to obtain the full geometric structure of images resulting
from strong gravitational lensing, we plan to pursue a similar analysis of the full Kerr bending angle beyond 
the equatorial plane.  Our goal is to study the dependence of the bending angle 
on the inclination and the resulting variations in the image positions and magnifications 
in a perturbative framework that without any assumptions about the inclination angle or the 
black hole spin.

\begin{acknowledgments}

S. V. I. thanks Arlie O. Petters for numerous helpful discussions.  E. C. H. was 
funded by the Dr. Jerry D. Reber Student/Faculty Research Fund at SUNY 
Geneseo.  The authors thank Kevin Cassidy for his support of this research fund.

\end{acknowledgments}

\end{document}